# Comment on "N-body simulations of gravitational dynamics" by Dehnen and Read (arXiv 1105.1082v1)

Adrian L. Melott, Dept. of Physics and Astronomy, University of Kansas (melott@ku.edu)

*Abstract:* It is shown that the historical summary of the growth in size of N-body simulations as measured by particle number in this review is missing some key milestones. Size matters, because particle number with appropriate force smoothing is a key method to suppress unwanted discreteness, so that the initial conditions and equations of motion are appropriate to growth by gravitational instability in a Poisson-Vlasov system appropriate to a Universe with dark matter. Published strong constraints on what can be done are not included in the review.

**TEXT**: Recently Dehnen and Read have posted a preprint[1] reviewing N-body simulations primarily in an astrophysical and cosmological gravitational context. They highlight the growth in the particle number in these simulations with a historical plot. However, two key milestones—the first published simulation results with $64^3$ particles[2] and $128^3$ particles[3] respectively—were not included.

I have reproduced below their Figure 1, with added points (in black) representing these two milestones, in order to set the record straight. These points lie considerably above the locus of blue points (intended to simulate collisionless systems) to which family they belong.

The emphasis on large numbers of particles is not an accident. It was obvious from the beginning that excessive discreteness and collisionality would spoil the fidelity of simulating dark matter gravitational clustering; some explicit tests are shown in (4). Tests continued for some time[5], which showed that although a group of cosmological simulations may satisfy a convergence test, they do not converge on a Vlasov-Poisson solution unless there is a mean global particle density of one per softening volume. These results were later confirmed in independent tests[6-7]. The reasons for this criterion are not difficult to understand, but as they imply severe limitations on what can be reliably simulated, these limitations are not often followed.

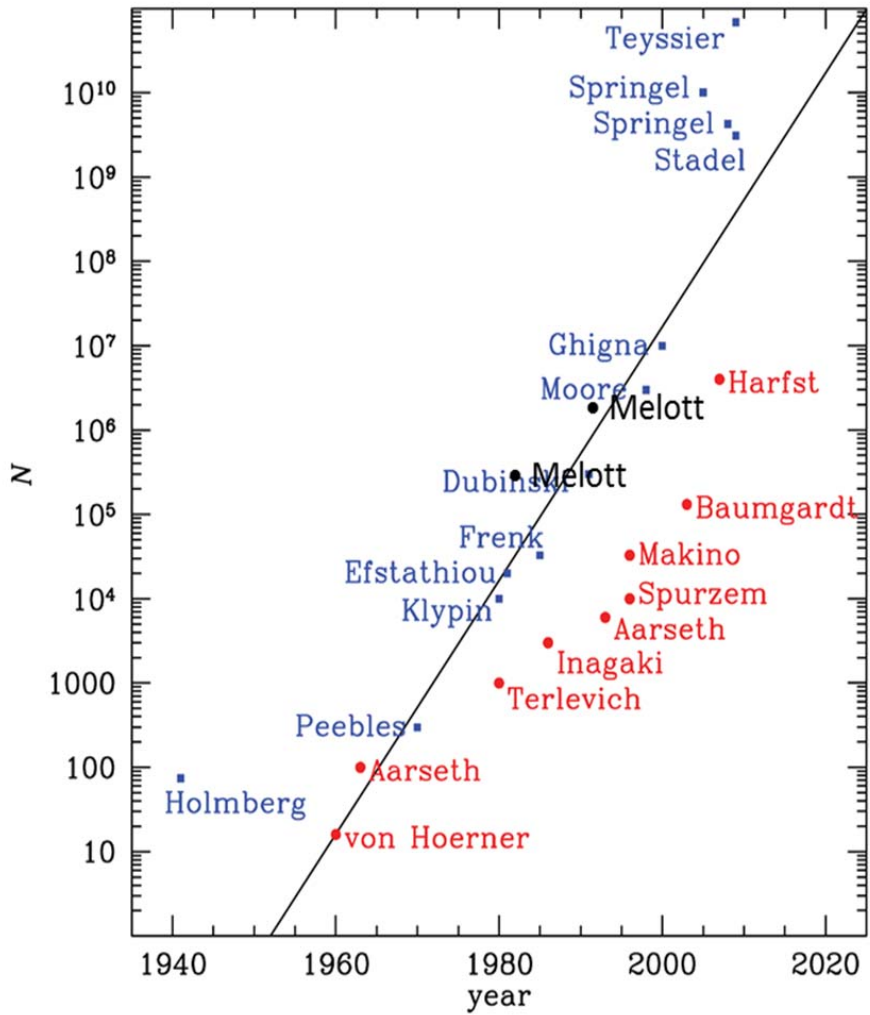

1. N-body simulations of gravitational dynamics (W. Dehnen and J. I. Read) arXiv:1105.1082
2. Three-Dimensional Simulation of Large Scale Structure in the Universe (J. Centrella and A. Melott) *Nature* **305**, 196 (1983).
3. Does Faint Galaxy Clustering Contradict Gravitational Instability? (A.L. Melott) *Astrophysical Journal Letters* **393**, 145 (1992).
4. A model for the formation of the Local Group (P.J.E. Peebles, A.L. Melott, M.R. Holmes, and L.R. Jiang) *Astrophysical Journal* **345**, 108 (1989).
5. Fundamental Discreteness Limitations of Cosmological N-Body Clustering Simulations (R.J. Splinter, A.L. Melott, S.F. Shandarin, and Y. Suto) *Astrophysical Journal* **497**, 38 (1998), and references therein.
6. Towards quantitative control on discreteness error in the non-linear regime of cosmological N-body simulations (M. Joyce, B. Marcos, and T. Baertschiger) *Monthly Notices of the Royal Astronomical Society* **394,** 751 (2009).
7. The Coyote Universe. I. Precision Determination of the Nonlinear Matter Power (K. Heitmann, M. White, C. Wagner, S. Habib, and D. Hignon) *Astrophysical Journal* **715**, 104 (2010).